\input{aipcheck}

\documentclass[
    ,final            % use final for the camera ready runs
  ]
  {aipproc}

\layoutstyle{8x11double}

\begin{document}

\title{Gravity Mediation in 6d Brane-World 
Supergravity \footnote{Talk given at PASCOS 2005, Gyeongju, 
Korea, May 30-June 4, 2005, and String Phenomenology 2005, M\"unchen, 
Germany, June 13-18, 2005.}}


\author{Hyun Min Lee}{
  address={Deutsches Elektronen-Synchrotron DESY, D-22603 Hamburg, Germany}
}

\begin{abstract}
We consider the gravity-mediated SUSY breaking
within the effective theory of six-dimensional brane-world supergravity.
We construct the supersymmetric
bulk-brane action by Noether method and find the nontrivial moduli coupling of
the brane F- and D-terms.
We find that
the low energy K\"ahler potential is not of sequestered form, so
gravity mediation may occur at tree level.
In moduli stabilization with anomaly effects included,
the scalar soft mass squared can be
positive at tree level and it can be comparable to the anomaly mediation.

\end{abstract}

\maketitle

%%%%%%%%%%%%%%%%%%%%%%%%%%%%%%%%%%%%%%%%%%%%
%% MAINMATTER
%%%%%%%%%%%%%%%%%%%%%%%%%%%%%%%%%%%%%%%%%%%%

\section{Introduction and summary}
When supersymmetry(SUSY) breaking is parametrized by soft mass
parameters which respect the Standard Model gauge symmetry, 
there would occur new dangerous CP and/or flavor violations. 
As a generic consequence of local SUSY,
gravity mediation of hidden sector SUSY breaking
generates all required soft mass terms of order
the weak scale \cite{gm}.
Generically, however, 
there appear contact terms between visible and hidden sectors
in the K\"ahler potential as
\begin{eqnarray}
\Omega&=&-3M^2_p+\Omega_0(Q_H,{\overline Q}_H) \nonumber \\
&&-\frac{1}{3M^2_p}\bigg(\delta_{ij}+C_{ij}(Q_H,{\overline Q}_H)\bigg)
{\overline Q}^i_VQ^j_V+\cdots.
\end{eqnarray}
For a non-diagonal contact term $C_{ij}$, the supergravity
F-term potential in the hidden sector  
would lead to visible soft scalar mass
with ${\cal O}(1)$ off-diagonal components.

In higher dimensional supergravity where extra dimensions are compactified
on orbifolds, one can find a more fundamental origin of contact 
terms \cite{randall}.
First, when the visible sector is located at the orbifold fixed point,
it can be geometrically separated from the hidden sector which is located 
at a different fixed point.
Second, while effects of heavy bulk fields to contact terms are suppressed 
for large extra dimensions, 
contact terms can be computed within the effective field theory 
by integrating out Kaluza-Klein modes of bulk fields. 
If SUSY breaking is mediated dominantly by higher dimensional gravity multiplet,
it is guaranteed that generated contact terms are flavor universal. 

For instance, in 
5d minimal supergravity compactified 
on $S^1/{\bf Z}_2$ \cite{lusu,ghri,bugago,rascst},
sequestering of visible sector is manifest in the low energy K\"ahler
potential ($K=-3\ln\Omega$) as
\begin{eqnarray}
\Omega=\frac{1}{2}(T+{\overline T})-\frac{1}{3}\Omega_V(Q_V,{\overline Q}_V)
-\frac{1}{3}\Omega_H(Q_H,{\overline Q}_H).
\end{eqnarray}
Because of the no-scale structure,
soft masses are zero at tree level, which is attributed to locality  
of SUSY breaking sector in extra dimension.
Therefore, in this case, the no-scale structure of tree-level
K\"ahler potential must be lifted by loop-corrections.
The anomaly mediation is a generic consequence in the off-shell formulation 
of 4d supergravity \cite{randall} 
and it may be a dominant contribution to soft masses. 
Since the anomaly mediation gives a flavor-universal contribution 
to soft masses, it would help amelierate the SUSY flavor problem. 
However, there arises a serious problem
with negative slepton mass squared.
Moreover, one-loop gravitational correction to 
the scalar soft mass squared also turned out to be negative \cite{rascst,aa}.
Generalizing the setup to warped supergravity does not improve 
the situation \cite{grrasc,aa}. 
In models with sizable brane gravity kinetic terms,
however, positive scalar masses can be obtained \cite{rascst,grrasc}.

On the other hand,
it has been pointed out in ref.~\cite{anisimov} that 
sequestering is absent in higher dimensional brane-world supergravity.
In this case, when KK modes of bulk fields 
are involved in the tree-level exchange between different sectors, 
their decoupling may lead to 
(generically non-diagonal) non-local contact terms at tree level 
which are not suppressed in comparison to the gravitino mass.

In this talk, we consider the gravity-mediated SUSY breaking 
in the context of 6d minimal supergravity \cite{masc,nise}. 
We compactify two extra dimensions on $T^2/{\bf Z}_2$ 
and impose the orbifold boundary conditions on the component fields 
of supergravity multiplet. We construct the supersymmetric 
bulk-brane action by Noether method and find the nontrivial moduli coupling of 
the brane F- and D-terms.  
From the obtained low energy effective action, we find that 
there is no sequestering in the K\"ahler potential any more. 
Because of a tree-level coupling between the visible and hidden
sectors, tree-level scalar soft masses 
in the visible sector may be generated. 
However, it crucially depends on the form of the effective superpotential.
If moduli are stabilized by the brane D-term 
with the effective superpotential which does not depend on the $T$ modulus,
there is a sequestering in the scalar potential.
On the other hand, when one takes into account the effect of anomalies, 
the superpotential gets a correction with the $T$ modulus dependence.
In this case, the scalar soft mass squared can be
positive at tree level and it can be comparable to the anomaly mediation.

\section{Supergravity on $T^2/Z_2$}
The 6d minimal supergravity multiplet is composed of
graviton ($e^A_M$), gravitino ($\psi_M$),
Kalb-Ramond field ($B_{MN}$), dilaton ($\phi$) and dilatino ($\chi$).
The 6d chirality conditions are 
$\Gamma^7\psi_M=\psi_M$ and $\Gamma^7\chi=-\chi$.
For the absence of 6d gravitational anomalies,
one needs additional multiplets with  
$n_H=244+n_V-n_T$
where $n_V={\rm dim G}$ and 
$n_H(n_T)$ is the number of hyper(tensor) multiplets.
However, we assume the additional multiplets get a heavy mass and they 
are irrelevant for SUSY breaking.
In this work, we focus only on the gravity mediation of SUSY breaking.

We regard two extra dimensions to be compactified
on a torus with periodicities 
$z\equiv z+2\pi(m+in\tau)R$
where $z=x_5+i\tau x_6$ with $x_{5,6}\equiv x_{5,6}+2\pi R$
and $\tau=\tau_2+i\tau_1$ is the complex structure.
In this coordinate basis, 
the 2d metric is given by $ds^2_2=g_{mn}dx^m dx^n$ with
\begin{eqnarray}
g_{mn}=-\frac{A}{\tau_2}\left[\begin{array}{ll} 1  & \,\,\,\,\,\,\,\, \tau_1 \\
\tau_1 & \tau^2_1+\tau^2_2 \end{array}\right].
\end{eqnarray}
Let us impose the ${\bf Z}_2$ orbifold symmetry on the torus. 
By identifying $(x_5,x_6)$ with $(-x_5,-x_6)$,
there appear four orbifold fixed points:
$(0,0),\ (\pi R,0), \ (0,\pi R), \ (\pi R, \pi R)$.
Decomposing the 6d spinor into 4d form as
$\Psi\equiv (\psi^1,{\overline {\psi^2}})^T$, we assign even $Z_2$-parity to
\begin{eqnarray}
e^{\bar \nu}_\mu, \ e^{\bar n}_m, \ B_{\mu\nu}, \ B_{mn}, \ \phi, \
\psi^1_\mu, \ \psi^2_m, \ \chi^2,
\end{eqnarray}
and odd $Z_2$-parity to
\begin{eqnarray}
e_{\mu {\bar m}}, \ B_{\mu m}, \ \psi^2_\mu, \ \psi^1_m, \ \chi^1.
\end{eqnarray}
Then, one massless ${\cal N}=1$ supergravity multiplet
is built from zero modes of $e^{\bar \nu}_\mu$
and $\psi^1_\mu$; three massless ${\cal N}=1$ chiral multiplets
from zero modes
of $e^{\bar n}_m,B_{\mu\nu},B_{mn},\phi,\psi^1_m$ and $\chi^2$;
and a KK tower of (short) massive multiplets of ${\cal N}=2$ supergravity
and two massive ${\cal N}=1$ vector multiplets.

\paragraph{Brane couplings of moduli}

A Noether construction of the bulk-brane action \cite{fll} gives 
a nontrivial dilaton coupling to a brane chiral multiplet as
\begin{eqnarray}
{\cal L}_Q=e_4\bigg[e^{-\phi}D^\mu Q^\dagger D_\mu Q
+\frac{i}{2}e^{-\phi}{\overline\psi}_Q\gamma^\mu D_\mu\psi_Q
+\cdots\bigg] 
\end{eqnarray}
Moreover, the brane action of a vector multiplet gives 
the brane D-term with a dilaton coupling as 
\begin{eqnarray}
{\cal L}_D=-e_4\frac{1}{2}e^{-2\phi}g^2D^2.\label{dterm}
\end{eqnarray}
In terms of ${\bf Z}_2$-even components of gravitino, 
we can construct a gravitino mass term at the fixed point as 
\begin{eqnarray}
{\cal L}_W&=&e_4\frac{1}{2}
\sqrt{\frac{A}{\tau_2}}W_0(-{\overline\psi}_\mu\gamma^{\mu\nu}
C{\overline\psi}^T_\nu
+i{\overline\psi}_+\gamma^\mu C{\overline\psi}^T_\mu \nonumber \\
&&+i{\overline\psi}_-\gamma^\mu C{\overline\psi}^T_\mu+{\rm h.c.})
\label{gravitino}
\end{eqnarray}
where $\psi_\pm=-(\psi^2_5\pm i\psi^2_6)$ and $W_0$ is a constant brane
superpotential. 
The result is generalized to a field-dependent brane superpotential $W=W(Q)$.
Thus, we also get the brane F-term with a nontrivial moduli dependence as
\begin{eqnarray}
{\cal L}_F=-e_4\frac{Ae^\phi}{\tau_2}|F|^2\label{fterm}
\end{eqnarray}
with $F=\partial_Q W$.

\section{Gravity mediation of supersymmetry breaking}

\paragraph{Low energy supergravity}

We take the 6d metric as
\begin{eqnarray}
ds^2=A^{-1}(x)g_{\mu\nu}(x)dx^\mu dx^\nu+ds^2_2
\end{eqnarray}
and plug the metric into the 6d action.
Then, from the resulting kinetic terms of bosonic action \cite{fll},
we find the K\"ahler potential in the 4d effective supergravity as
\begin{eqnarray}
K&=&-{\rm log}\bigg(\frac{1}{2}(T+{\overline T})-\frac{1}{M^2_p}|Q|^2\bigg)
\nonumber \\
&&-{\rm log}\bigg(\frac{1}{2}(S+{\overline S})\bigg)
-{\rm log}\bigg(\frac{1}{2}(\tau+{\overline \tau})\bigg)
\end{eqnarray}
with $M^2_p\equiv (2\pi)^2M^4_6$, and
\begin{eqnarray}
T&=&t+\frac{1}{M^2_p}|Q|^2+i\sqrt{2}B_{56}, \nonumber \\
S&=&s+i\sqrt{2}a, \ \ \ \tau=\tau_2+i\tau_1, \label{moduli}
\end{eqnarray}
where $s=Ae^{-\phi},t=Ae^{\phi}$, and
$e^{-2\phi}H_{\mu\nu\rho}=\epsilon_{\mu\nu\rho\sigma}\partial^\sigma a$.
Moreover, we find that the brane gauge kinetic function 
does not depend on moduli.
By comparing eq.~(\ref{gravitino}) to the 4d standard formula for the gravitino
mass as 
${\cal L}_m=-\frac{1}{2}e_4 e^{K/2} W{\bar\psi}_\mu\gamma^{\mu\nu}C{\bar\psi}^T_\nu$, we can also see that 
the effective brane superpotential does not depend on moduli.

\paragraph{Gravity mediation}
With the visible and hidden sectors at two different
fixed points, the relevant Lagrangian in 4d superconformal frame is given by
\begin{eqnarray}
{\cal L}=-3M^2_P\int d^4\theta\Phi^\dagger\Phi\Omega
\end{eqnarray}
where $\Phi=1+\theta^2F_\Phi$ is the compensator field,
and $\Omega=e^{-K/3}$ is given by
\begin{eqnarray}
\Omega&=&\frac{1}{2}(T+{\overline T}-2\Omega_V-2\Omega_H)^{1/3}
(S+{\overline S})^{1/3} (\tau+{\overline \tau})^{1/3} \nonumber \\
&\simeq&\frac{1}{2}(T+{\overline T})^{1/3}(S+{\overline S})^{1/3}
(\tau+{\overline \tau})^{1/3}\nonumber \\
&&\times\bigg[1-\frac{2}{3}
\frac{\Omega_V+\Omega_H}{T+{\overline T}}
-\frac{8}{9}\frac{\Omega_V\Omega_H}{(T+{\overline T})^2}\bigg].
\end{eqnarray}
As a consequence, we find  
that there is no sequestering of visible sector 
in the K\"ahler potential even at tree level.
For general SUSY breaking including the brane D-term, 
scalar soft mass in the visible sector is expressed in terms of 
the gravitino mass $m_{3/2}$ and F-terms as
\begin{eqnarray}
m^2_{Q_V}&=&4m^2_{3/2}-2\frac{|F_T|^2}{(T+{\overline T})^2}
-\frac{|F_S|^2}{(S+{\overline S})^2} \nonumber \\
&&-\frac{|F_\tau|^2}{(\tau+{\overline \tau})^2}
-4\frac{|F_H|^2}{(T+{\overline T})},
\end{eqnarray}
with $F_i=-e^{K/2}(K^{-1})_{ij}D_jW$,
while a vanishing cosmological constant imposes a condition 
\begin{eqnarray}
\frac{2D^2}{M^2_P(T+{\overline T})^2}
&=&3m^2_{3/2}-\frac{|F_T|^2}{(T+{\overline T})^2}
-\frac{|F_S|^2}{(S+{\overline S})^2} \nonumber \\
&&-\frac{|F_\tau|^2}{(\tau+{\overline \tau})^2}
-2\frac{|F_H|^2}{(T+{\overline T})}.
\end{eqnarray}

\section{Moduli stabilization and soft mass terms}

From the brane terms (\ref{dterm}) and (\ref{fterm}) with eq.~(\ref{moduli}), 
we obtain the effective scalar potential as 
\begin{eqnarray}
V=\frac{|F|^2}{[(S+{\overline S})/2][(\tau+{\overline\tau})/2]}
+\frac{D^2}{[(T+{\overline T})/2-|Q|^2]^2}.
\end{eqnarray}
This is of the run-away type, so there is no local minimum for moduli.
Thus, in order to stabilize moduli, 
we need additional contributions in the scalar potential due to bulk dynamics.

\paragraph{Bulk gaugino condensation}
The gauge kinetic function for zero mode of bulk vector multiplet is,
at tree level,
\begin{eqnarray}
f=\frac{1}{g^2_4}S
\end{eqnarray}
where $g^2_4=g^2_6/(2\pi R)^2$ with $g_6$ the 6d gauge coupling.
When there are two gaugino condensates in the bulk below the compactification
scale, the nonperturbative effective superpotential  
is generated to be of racetrack form \cite{racetrack},
\begin{eqnarray}
W=(\eta(i\tau))^{-2}(\Lambda_1 e^{-c_1S}
+\Lambda_2e^{-c_2S})
\end{eqnarray}
where $\eta(i\tau)$ is the Dedekind eta function,
$c_1, c_2,\Lambda_1,\Lambda_2$ are constants.
Here we note that the moduli invariance associated with the shape
of a torus requires the $\tau$ dependence in the superpotential. 

\paragraph{Stabilization of $S$ and $\tau$ }
In the presence of the nonperturbative superpotential,
the scalar potential is minimized for $F_S=F_\tau=0$.
Then, the local minimum of $S$ is
\begin{eqnarray}
{\rm Re}S&=&\frac{1}{c_1-c_2}
{\rm log}\bigg[\frac{\Lambda_1(2c_1{\rm Re}S+1)}
{\Lambda_2(2c_2{\rm Re}S+1)}\bigg], \\
{\rm Im}S&=&\frac{\pi(2n+1)}{c_1-c_2}, \
n\in {\bf Z},
\end{eqnarray}
and
\begin{eqnarray}
{\hat G}_2\equiv -2\pi\bigg(\partial_\tau{\rm log}\eta(i\tau)
+\frac{1}{\tau+{\overline\tau}}\bigg)=0,
\end{eqnarray}
leads to a local minimum $\tau=\sqrt{3}/2+i/2$
(another zero $\tau=1$ is a saddle point.).

\paragraph{Stabilization of $T$}
After stabilization of $S$ and $\tau$, 
there remains the scalar potential for $T$ as
\begin{eqnarray}
V/M^2_p&=&-\frac{2{\hat m}^2_{3/2}-|{\hat F}_S|^2-|{\hat F}_\tau|^2}
{(T+{\overline T})}+2|{\hat F}_H|^2 \nonumber \\
&&+\frac{2D^2}{(T+{\overline T})^2}
\end{eqnarray}
where ${\hat m}_{3/2},{\hat F}_S,{\hat F}_\tau$ and
${\hat F}_H$ are $T$-independent constants proportional to the quantities
without the hat.
The $T$-modulus is stabilized with zero vacuum energy,
provided that
\begin{eqnarray}
\frac{2D^2}{(T+{\overline T})^2}=2|{\hat F}_H|^2
=\frac{{\hat m}^2_{3/2}-\frac{1}{2}|{\hat F}_S|^2-\frac{1}{2}|{\hat F}_\tau|^2}
{(T+{\overline T})}.
\end{eqnarray}
After stabilization of all moduli,
the scalar soft mass vanishes.
One-loop gravity correction to the K\"ahler potential does
not generate a soft mass either.
The real part of $T$-modulus gets a mass as
\begin{eqnarray}
m^2_T=\frac{8D^2}{(T+{\overline T})^2}\simeq 4m^2_{3/2}.
\end{eqnarray}
The axionic part of $T$-modulus could be eaten up by a bulk gauge field
via Green-Schwarz mechanism.

\paragraph{Anomaly corrections}
Even though the 6d irreducible anomalies can be cancelled by the introduction
of additional multiplets, generically one also 
needs a Green-Schwarz counterterm \cite{gs}
to cancel the reducible part of anomalies.
Then, after a supersymmetric completion of the GS term \cite{sag,gmw}, 
the bulk gauge kinetic term gets modified to 
\begin{eqnarray}
f_a=\frac{1}{g^2_a}(S+\alpha_a T)
\end{eqnarray}
where $\alpha_a$ depends on bulk fermion content.
As illustation, when there are multiple gaugino condensates 
with anomaly corrections, the effective superpotential can be 
\begin{eqnarray}
W=(\eta(i\tau))^{-2}W(S)W(T)
\end{eqnarray}
with
\begin{eqnarray}
W(S)&=&\Lambda_1 e^{-c_1S}+\Lambda_2 e^{-c_2S}, \\
W(T)&=&\Lambda_3 e^{-c_3T}+\Lambda_4 e^{-c_4T}.
\end{eqnarray}
For $F_S=F_\tau=F_T=0$, there exists a AdS minimum with all moduli stabilized.
Then, the $T$-modulus mass is given by
\begin{eqnarray}
m^2_T\sim c^2_3c^2_4(T+{\overline T})^4 m^2_{3/2}.
\end{eqnarray}
For $|c_3-c_4|\ll c_3$, $(T+{\overline T})$ can be large enough
and $m^2_T\gg m^2_{3/2}$.

In order to lift the vacuum up to a Minkowski vaccuum, one needs 
a positive lifting potential such as the brane F-term 
and/or D-term. In this lifting procedure, SUSY is broken and a nonzero soft 
mass can be generated at tree level.
When the D-term lifting potential is dominant as  
\begin{eqnarray}
\frac{D^2}{(T+{\overline T})^2}\sim m^2_{3/2}M^2_P,  \ \
\frac{|F_{H}|^2}{(T+{\overline T})}\ll m^2_{3/2},
\end{eqnarray}
the scalar soft mass squared tends to be positive.
If the D-term is comparable to the F-term, 
tree-level soft masses can be comparable 
to the contribution from anomaly mediation.

\begin{theacknowledgments}
The author thanks A.~Falkowski and C.~L\"udeling 
for a nice discussion during the collaboration.
\end{theacknowledgments}



\end{document}